\documentstyle[preprint,aps,epsfig]{revtex}

\begin{document}

\title{Evidence for Three Nucleon Force Effects in p-d Elastic
Scattering}\author{A. Kievsky$^{1,2}$
M.H. Wood$^2$, C.R. Brune$^2$, B.M. Fisher$^2$, H.J. Karwowski$^2$,
D.S. Leonard$^2$, E.J. Ludwig$^2$, S. Rosati$^{1,3}$, M. Viviani$^1$}

\address{{$^1$}Istituto Nazionale di Fisica Nucleare, 
         Via Buonarroti 2, 56100 Pisa, Italy}
\address{{$^2$} Department of Physiscs and Astronomy, University of
         North Carolina at Chapel Hill, Chapel Hill, NC 27599-3255, USA
         and Triangle Universities Nuclear Laboratory, Durham, NC 27708, USA}
\address{{$^3$}Department of Physics, University of Pisa, 56100 Pisa, Italy}

\maketitle

\begin{abstract}
A new measurement of the p-d differential cross section at
$E_p=1$ MeV has been performed. These new data and older
data sets at energies below the deuteron breakup are compared to 
calculations using the two--nucleon Argonne $v_{18}$
and the three--nucleon Urbana IX potentials. A quantitative
estimate of the capability of these interactions to describe the
data is given in terms of a $\chi^2$ analysis. The $\chi^2$
per datum drastically improves when the three-nucleon interaction
is included in the Hamiltonian.  
\end{abstract}

\newpage

The new generation of NN potentials describes the
two-nucleon (2N) observables with a $\chi^2$ per datum 
$\approx 1$~\cite{NIJM,AV18,CDBONN}. 
This  high accuracy obtained in the description of the 2N
system does not imply that a similar accuracy will be
achieved in the description of larger nuclear systems, in particular
the three-nucleon (3N) data. In fact, the
simplest observable in the 3N system, the binding energy, is
underpredicted by each  of the new NN potentials. The
energy deficit ranges from $0.5$ to $0.9$ MeV depending
on  the off-shell and short range parametrization
of the NN interaction. This underbinding problem
has not yet been solved, and a number of effects beyond the
static NN interaction have been considered ( a review is given 
in ref.~\cite{CS98}). For example, considerable efforts have been put
into calculating relativistic corrections and three-nucleon force 
(3NF) contributions to the 3N binding energy.

It is common practice to 
look at the 3N bound state problem as the solution of the non-relativistic
Schr\"odinger equation using phenomenological NN interactions and
then to introduce a 3NF to provide supplementary binding. 
The models for the 3NF are usually based on two--pion 
exchange with intermediate $\Delta$-isobar excitation,
and the  strength of the interaction is adjusted  to reproduce the
${}^3$H binding energy.

Once the 3N binding energy is well reproduced, the description of
several other observables improves as well. For example, the $A=3$ r.m.s 
radii~\cite{radii}, the asymptotic normalization constants
$\eta$~\cite{kb97}
and the doublet n-d scattering lengths~\cite{KVR95} are now in much
better agreement with the experimental values. These observables 
have the property to scale with 3N binding energy (the so-called
Phillips lines)~\cite{phillips}. 

With respect to the 3N continuum, a complete quantitative 
analysis in terms of $\chi^2$ of the 3N data versus theory
has not yet been made for any of the new NN potentials. 
Therefore, there is a need to evaluate  in detail the 
ability of those interactions to describe the 3N scattering data. 
In ref.~\cite{report} a detailed analysis has been 
performed for the total
n-d cross section in which calculations solving the
Faddeev equations have been compared to the data. 
This analysis has been recently repeated ~\cite{WKN99} by
taking into account  new high-precision measurements~\cite{ABD98}.
The analysis could not be extended to the differential 
cross section, due to lack of an  adequate data set. In ref.~\cite{sakai}
a new set of precise measurements of d-p elastic observables at 
$E_d=270$ MeV has been presented. The differential cross section as
well as some polarization observables has been analyzed with Faddeev
calculations
using modern NN potentials including 3NF contributions. The $\chi^2$ per
datum has been studied in a limited angular range
($\theta_{c.m.}=50^\circ - 180^\circ$) in order to avoid the
effects of the Coulomb interaction, which has been neglected in that
calculations.
At this very high energy a definite sensibility to 
three-body forces has been observed.

Recently a rigorous solution of the p-d scattering 
problem has been obtained by the Pisa group~\cite{KVR95,KVR94}
allowing for a detailed study of this reaction for which an extensive 
and high precision data set exists.
In refs.~\cite{KLA93,KRTV96} phase shift analyses have been performed
in order to reproduce the p-d differential cross section and vector and
tensor analyzing powers. From these analyses it was possible to make
comparisons to the theoretical phase-shift and mixing parameters 
and quantitatively relate
the found differences in the $P$--wave parameters to the so called
"$A_y$ puzzle"~\cite{KRTV96}.

In the present paper we use these calculations in an attempt to
analyze in a quantitative way the capability of the
modern NN interactions to describe the p-d differential cross section
in the low energy regime.
To this aim we will present a new precise measurement of the p-d
differential cross section at $E_{p}=1$ MeV and its theoretical
description in terms of the Argonne $v_{18}$ potential (AV18)~\cite{AV18}
plus the Urbana 3NF (UR)~\cite{UR}.
The new high precision data were taken as a part of a measurement program 
of p-d scattering observables including the two vector
($A_y$ and $iT_{11}$) and the three tensor
($T_{20}$, $T_{21}$, and $T_{22}$) analyzing powers at $E_{p}=1$ MeV. 
The complete set of data will be published
elsewhere~\cite{wood00} and is part of a program developed at
Triangle Universities Nuclear Laboratory (TUNL)
to study the properties of the few body interactions at low energies.

The measurement of  cross sections for p-d scattering was
conducted at TUNL using the 10 MV FN tandem accelerator. The 
deuteron beam   was accelerated to an
energy of 2.0~MeV and directed by a dipole magnet to a
scattering chamber. The magnet and a feedback 
system with the FN tandem kept
the beam energy constant to within  $\pm5$~keV.  
The targets were made of thin hydrogenated
carbon foils containing approximately
$0.5\times10^{18}$~H/cm$^2$ and $1.0\times10^{18}$~C/cm$^2$~\cite{tim}
and were replaced  often during the experiment.
The elastically-scattered deuterons and recoil 
protons were counted in two pairs of silicon surface-barrier detectors 
placed  $10^\circ$ apart symmetrically with respect to the
beam direction.  In the measurement of relative cross
sections the detectors covered an angular range
of $\theta_{lab}=7^\circ$ to $64^\circ$ ($\theta_{c.m.}=21^\circ$ to
$166^\circ$).  Another two pairs of silicon
detectors were mounted to the chamber wall and set at $15^\circ$
and $42^\circ$  to normalize the yields in
the rotating detectors. The statistical accuracy of the
relative cross sections was less than $0.5$\% and the
systematic error was less than $0.8$\%.  A sample spectrum for
${}^1{\rm H}(d,d)$ scattering at $\theta_{lab}=26^\circ$ is shown in
Fig.1.

The absolute normalization of the cross section measurement
was obtained by relating the
measured differential p-d cross sections to the
well-known p-p cross section~\cite{Nim}.
The normalization procedure involved producing in sequence 
proton and deuteron beams of the same magnetic rigidity.
This procedure assured that both beams were
transported through the
beamline in the same way  and  could  be put on
approximately the same spot on the target by only adjusting
the dipole magnet after the ion source and changing the FN tandem
terminal voltage.
The scattering chamber was left with
the same setup as for
the relative measurement with detectors placed at two angles
which provided three  normalization  points [${}^1{\rm
H}(d,d)$ and ${}^1{\rm H}(d,p)$ for $25.0^\circ$ and ${}^1{\rm
H}(d,p)$ for $35.0^\circ$].
Each target used in normalization runs  remained in the beam
for a very short period of  time ($\approx$ 10 $\mu$C) to
reduce the effects of target deterioration. After the proton 
beam was put on target,
three targets were cycled as in the case of the measurements
with the deuteron beam. This
process of switching from deuteron beam to proton beam back to
deuteron beam was repeated four times with consistent results. 

An  Au target was utilized to determine   systematic
errors in the experimental setup and data collection process. 
At incident energies of
$E_{d}=2.0$~MeV and $E_{p}=4.0$~MeV, the ${}^{197}{\rm
Au}(d,d)$ and ${}^{197}{\rm Au}(p,p)$
cross sections should follow the Rutherford formula.
For our tests, the detectors pairs were placed at $140.0^\circ$ and
$150.0^\circ$. The target consisted of  170~$\mu$g/cm$^2$ of
Au evaporated on a 10~$\mu$g/cm$^2$ carbon foil.  At the
end of each cycle of runs with hydrogenated carbon targets, the Au target
was placed in the beam (either deuteron or proton).
The ratio of the ${}^{197}{\rm Au}(d,d)$ and ${}^{197}{\rm
Au}(p,p)$ scattering measurements were found to be within
0.6\% of the calculated values. 
The overall accuracy of the present absolute cross section
measurements is $0.8$\%.

The calculations of p-d scattering have been done
using the Pair Correlated Harmonic basis ~\cite{KVR94} to expand
the scattering p-d wave function.  The scattering matrix
has been obtained using the Kohn variational principle in its
complex form~\cite{kiev97} and, successively,
the cross section has been calculated using the
formula given in eq.(4) of ref.\cite{Seyler}. The accuracy 
of this method in the
calculations of the phase shift and mixing parameters
has been studied in ref.~\cite{kiev97}. In ref.~\cite{kiev98} 
a detailed comparison has been performed by comparing the present
technique with the results obtained by solving the Faddeev equations 
in momentum space. The numerical accuracy of the present technique
has been found to be of the order of $0.1$\%.

The results of the measurements
 for the p-d cross section at $E_{lab}=1$ MeV
are given in Fig. 2 (open circles) and compared to
the theoretical predictions.  The two curves shown in Fig. 2
correspond to calculations using the AV18 potential (dotted line)
and including also the Urbana 3NF (solid line).

There is a good agreement between the scattering data and
both calculations, though the cross section calculated using the 
AV18 potential is slightly higher than the one obtained with 
the AV18+UR model. This can be understood as arising 
from the additional  attraction
introduced by the 3NF which overall increases the binding of $^3$He and,
at low energy, reduces the cross section.
It is known that the inclusion of the Urbana 3NF modifies 
mainly the $J=1/2^+$ state, which is reflected in a change 
in the ${}^2S_{1/2}$ phase shift and the $\eta_{1/2+}$ mixing parameter
as noted in ref.~\cite{KLA93}.  
It is also evident from the figure
that the calculation using the AV18+UR potential, which gives a better
description of the bound system, also gives better agreement with the
elastic scattering data.

In order to gain a better understanding of the quality of the
agreement between theory and data, in Fig. 3 we present
the values of experimental differential cross section data divided by the
theoretical values calculated with  the AV18+UR potential model.
In this plot we also present the results of two other measurements
of differential cross sections at the same  energy~\cite{clegg,huttel}. 
The present data (open circles)
and the theoretical predictions are in agreement within $1$\%. This is
also the case with the data from ref.~\cite{clegg} (open squares)
though the theoretical cross section seems to be below these data.
We observe disagreement of $1-2$\% between our data and data 
of ref.~\cite{huttel} (open triangles) at forward angles.
These data are slightly below the theoretical cross section at forward
angles and slightly above
at backward angles. This behavior is in general present
when the comparison between theory and experiment is performed
at somewhat different energies.

A $\chi^2$ per datum analysis of the
theoretical calculations with respect to the experimental data is made
by the evaluation of the quantity

\begin{equation}
\chi^2={1\over N}\sum_i{(cf^{exp}_i-f^{th}_i)^2 \over (\Delta f_i)^2} \ ,
\end{equation}
where $f^{exp}_i$ is the $i$--th datum  at angle $\theta_i$
and $\Delta f_i$ its error, $f^{th}_i$ is the theoretical
value at the same angle and the total number of points is $N$.
The parameter $c$ is introduced to allow a variation in the absolute
normalization of the data. Its value is slightly varied around $c=1$
looking for a minimum in the value of $\chi^2$.
This is illustrated in Fig.4(a) where 
$\chi^2$ has been plotted as a function of the parameter
$c$ comparing the present data ($N=56$) 
to the AV18+UR theoretical cross section (solid line).
A minimum has been obtained
at the value $\chi^2=1.03$ by lowering
the normalization of the data by $0.2$\%. A similar
analysis using the data from ref.~\cite{clegg} ($N=12$) gives
a minimum at $\chi^2=0.26$ by lowering the data by $1.0$\%, though a
change of $0.6$\% is enough to obtain a $\chi^2=1$.
In both cases the change in the absolute normalization
is within the limits defined by the systematic errors. 
Therefore we can conclude that the calculation 
of the differential cross sections using 
the AV18+UR potential model gives
a $\chi^2$ per datum $\approx$ 1.0. 
The analysis of the data from ref.~\cite{huttel} ($N=20$) gives
a value of $\chi^2\approx 6$.

It is instructive to perform the   $\chi^2$ 
analysis of the present data with the calculation using the AV18
potential even
though it underbinds $^3{\rm He}$ by 0.8~MeV. In this case, with $c=1$,
 the result is $\chi^2\approx50$. This value can be
reduced to $\chi^2\approx8$ by changing the normalization by $3$\%,
which is far outside systematic 
errors of the data  and is equal to 28 when restricted to the limits of
the quoted systematic error. Therefore, a
second conclusion can be reached that the differential cross section can
not be correctly described using one of the new modern NN interactions,
in this case the AV18 potential.  Following the studies of the 
Bochum-Cracow group on the sensitivity of the n-d differential 
cross sections  to the different potentials (see ref.~\cite{report} 
page 163), this  conclusion should be valid for
the other modern NN interactions as well.
Thus, we have observed large three-nucleon force effects
in the p-d differential cross section at low energies
through a detailed $\chi^2$ analysis between theory and data.

The same analysis can be performed at other energies. 
Here we will limit the analysis to energies below the deuteron
breakup threshold in order to avoid the appearance of open channels.
High quality measurements exist at $E_{lab}=2.0$ and $3.0$ 
MeV~\cite{clegg,sagara}. These data are compared to the
cross sections  predicted with AV18+UR potential in Fig. 5.
In order to establish the quality of the agreement between theory
and experiment, the calculations of $\chi^2$ are
given in Fig. 4(b,c) for the two sets of measurements at both energies.
At $2$ MeV, the two data points of ref.~\cite{sagara} 
at most forward angles 
disagree distinctly with  the other data and 
are not included in the data base. 
Analysis of these data gives the value of
$\chi^2<1$ which is obtained by changing the total normalization less than
$1$\% (solid line) at both energies.  
For the data set of ref.~\cite{clegg} at $E_{lab}=2$ MeV
the minimum is at $\chi^2=2.9$ with a change in the
absolute normalization of $0.3$\%. At $E_{lab}=3$ MeV
the value $\chi^2=2.4$ is obtained and it can be reduced 
to $\chi^2=1.8$ when the outlier point at $158^\circ$ is removed. 

In Table I we collect the $\chi^2$--values obtained from the analysis
of the cross section data at three energies. 
For the calculations with AV18+UR potential  the $\chi^2$ values obtained
with  $c=1$ as well as the minimum $\chi^2$ found by  
varying the parameter $c$ are given. Remarkably, the present data
and the high quality data of ref.~\cite{sagara} at $E_{lab}=2$
and $3$ MeV give a $\chi^2 \le 1$
allowing less than $1$\% variation  in the absolute normalization of the
data. For the sake of comparison,
calculations using the AV18 potential are also given in Table I.
The $\chi^2$ using the values of $c$ previously optimized 
for AV18+UR potential as well as the minimum
$\chi^2$ obtained after the variation of the parameter $c$ are shown.
The minimization procedure improves the $\chi^2$ by a
factor of 5 to 10. However  the change in the absolute 
normalization is about $3$\%, considerably outside the limit due
to systematic errors of the data. Considering the 
present data and the data of ref.~\cite{sagara} we observe
that the $\chi^2$ for the calculations with AV18 potential
decreases as the energy increases. This trend could be a manifestation
of the previously observed fact that the cross section is
overpredicted at low energies and drops below the data for energies above
$30$ MeV~\cite{report}.

In conclusion we have presented a new high-quality measurement
of the p-d differential cross section at $E_p=1$ MeV with
absolute normalization to p-p elastic scattering.
The new measurement allows for a detailed comparison of the data 
to the calculated cross section using one of the 
new NN forces, the AV18 interaction,
with and without the inclusion of the Urbana 3NF.
In addition, the present data help to resolve a
significant descrepancy which existed between previous experiments
performed at this energy.
The use of $\chi^2$ analysis for these comparisons provides a quantitative
measure of the ability of various
NN and 3N Hamiltonians to reproduce the experimental data.
The calculations
with the AV18+UR potential are in excellent agreement with the
data with a $\chi^2$  per datum
$\approx 1$. 
The same degree of agreement is obtained with respect to
the data from ref.~\cite{sagara} at $E_p=2$ and $3$ MeV. 
However, at these two energies the data from 
ref.~\cite{clegg} show some scatter and the $\chi^2$ 
does not reach unity.

The calculations using only the AV18 potential give much  larger
values of $\chi^2$. This inability  of AV18 potential to  describe
adequately 
the 3N scattering data confirms the evidence of a deficiency
of modern NN potentials which need to be supplemented 
with a 3NF  to avoid the underprediction of the binding
energy of $^3{\rm He}$.
In order to provide evidence of 3NF effects beyond those
related to the correct description of the binding energy the present
analysis of the differential cross section has to be extended to higher
energies and to other observables such as the vector and tensor analyzing 
powers. Studies along these
lines have already begun~\cite{brune98,kiev99} and are presently being
pursued vigorously.

\begin{acknowledgements}
 
 One of the authors (A. K.) would like to thank the Triangle Universities 
 Nuclear Laboratory for hospitality and support during the 
 realization of this work. The authors would like to thank T.B. Clegg
 for useful discussions. This work was sponsored in part by the U.S.
Department of Energy, Grant No. DOE-FG02-97ER41041.
\end{acknowledgements}

\begin{table}
\caption{ $\chi^2$ per datum of the AV18 and AV18+UR p-d differential 
cross section compared to the present data and to the
data from ref.[23,25] at three different energies. The number
in parenthesis corresponds to the value of the parameter $c$ defined in
eq.(1).}
\begin{tabular}{c|cc|cc|cc}
\hline
         & \multicolumn{2}{c} {$E_p=1$ MeV} & 
           \multicolumn{2}{c} {$E_p=2$ MeV} & 
           \multicolumn{2}{c} {$E_p=3$ MeV} \\
         & present & ref.[23] & ref.[25] & ref.[23] & ref.[25] & ref.[23] \\ 
\hline
 AV18+UR & 1.15 (1.) & 3.43 (1.) & 1.01 (1. )& 3.34 (1.) & 
          3.24 (1.) & 4.52 (1.) \\
        & 1.03 (0.998) & 0.26 (0.990) & 0.53 (0.995 )& 2.97 (1.004) & 
          0.89 (1.010) & 1.80 (1.010) \\
\hline
 AV18 & 50.2 (0.998) & 22.7 (0.990) & 16.9 (0.995 )& 24.5 (1.004) & 
        15.8 (1.010) & 13.8 (1.010) \\
      & 7.66 (1.030) & 3.70 (1.020) & 2.09 (1.026 )& 5.06 (1.030) & 
        1.28 (1.038) & 2.92 (1.032) \\
\hline
\end{tabular}
\label{tb:chi2}
\end{table}

\newpage

\centerline{FIGURES}

Figure 1.  Typical spectrum of particles resulting from 
scattering the deuteron beam on thin hydrogenated 
carbon foil from relative cross section experiment.

Figure 2. Present data (open circles) for the p-d differential
cross section are compared to the theoretical curves calculated
with the AV18 potential (dotted line) and the AV18+UR potential
(solid line).

Figure 3. Present data (open circles) and the data from ref.~\cite{clegg} 
(open squares) and from ref.~\cite{huttel} (open triangles) divided
by the values calculated using  AV18+UR potential.

Figure 4. $\chi^2$ per datum as a function of the absolute normalization
at $E_p=1$ MeV (a), $2$ MeV (b) and
$3$ MeV (c) obtained by comparing the cross sections calculated
using AV18+UR potential to:
(a) present data (solid line) and data from ref.~\cite{clegg} (dotted line), 
(b) and (c) data from ref.~\cite{sagara} (solid line) and ref.~\cite{clegg}
(dotted line).

Figure 5. The p-d differential cross section calculated using AV18+UR
potential (solid line)
 compared to the data from ref.~\cite{clegg} (open squares) and 
ref.~\cite{sagara} (open circles) at $2$ MeV (a) and at $3$ MeV (b).

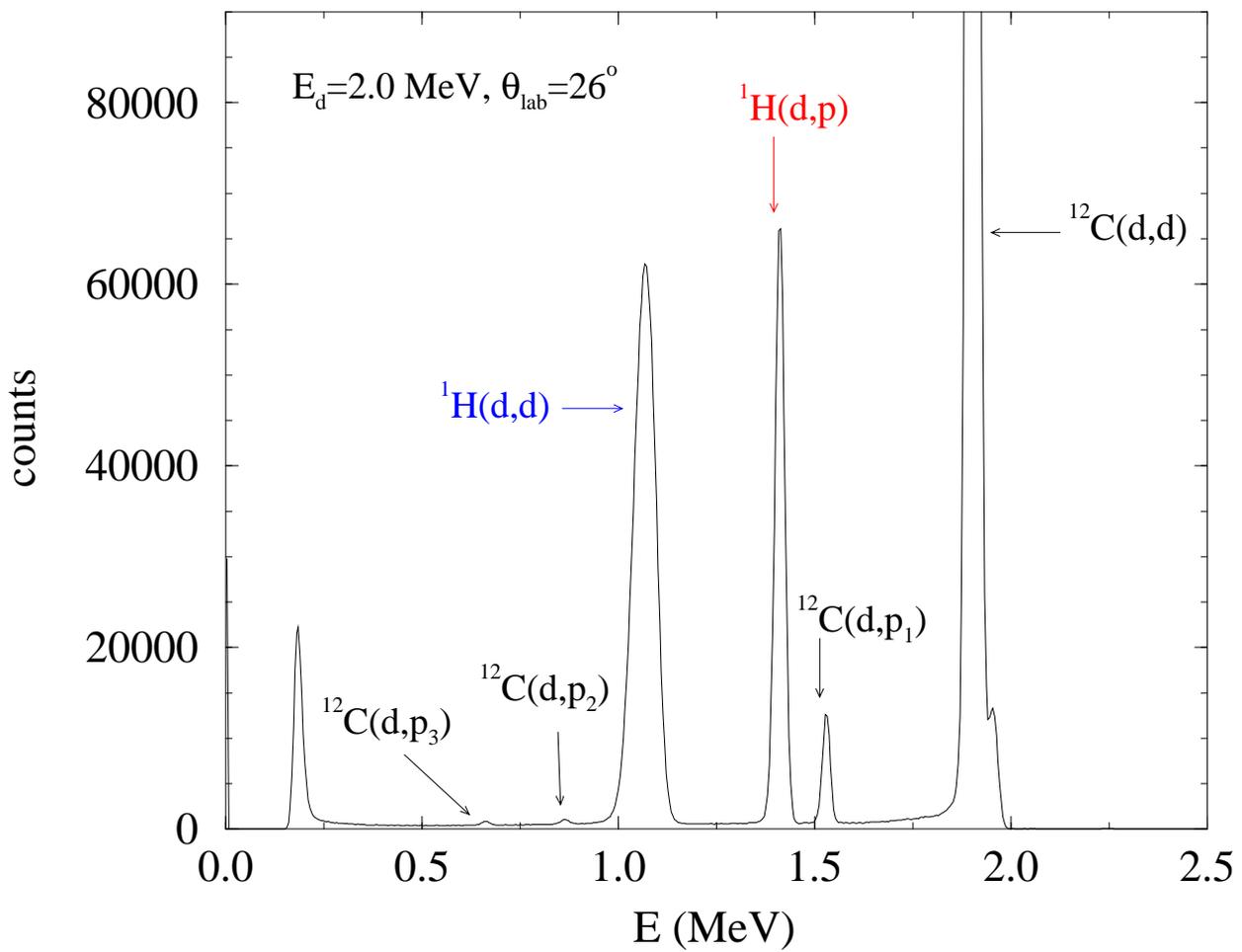
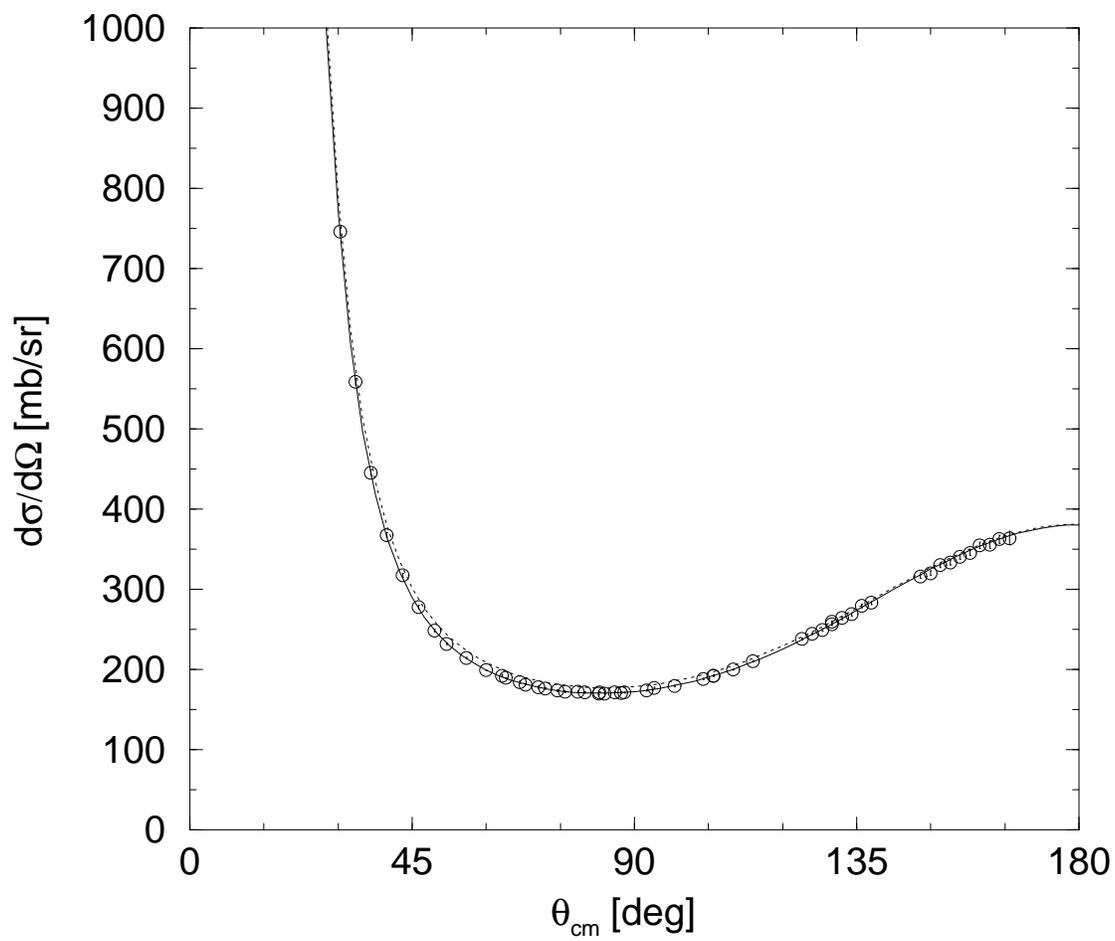
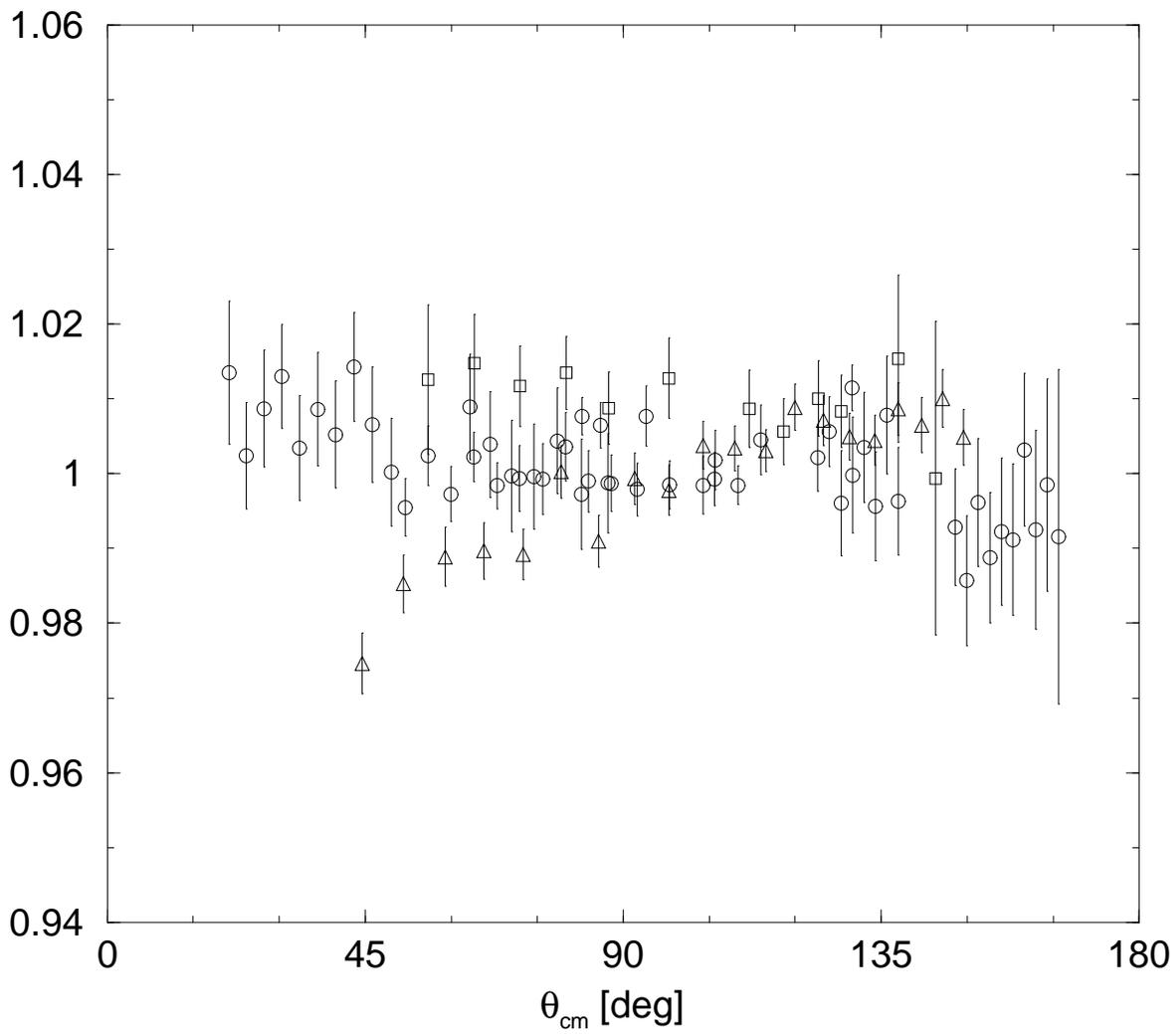
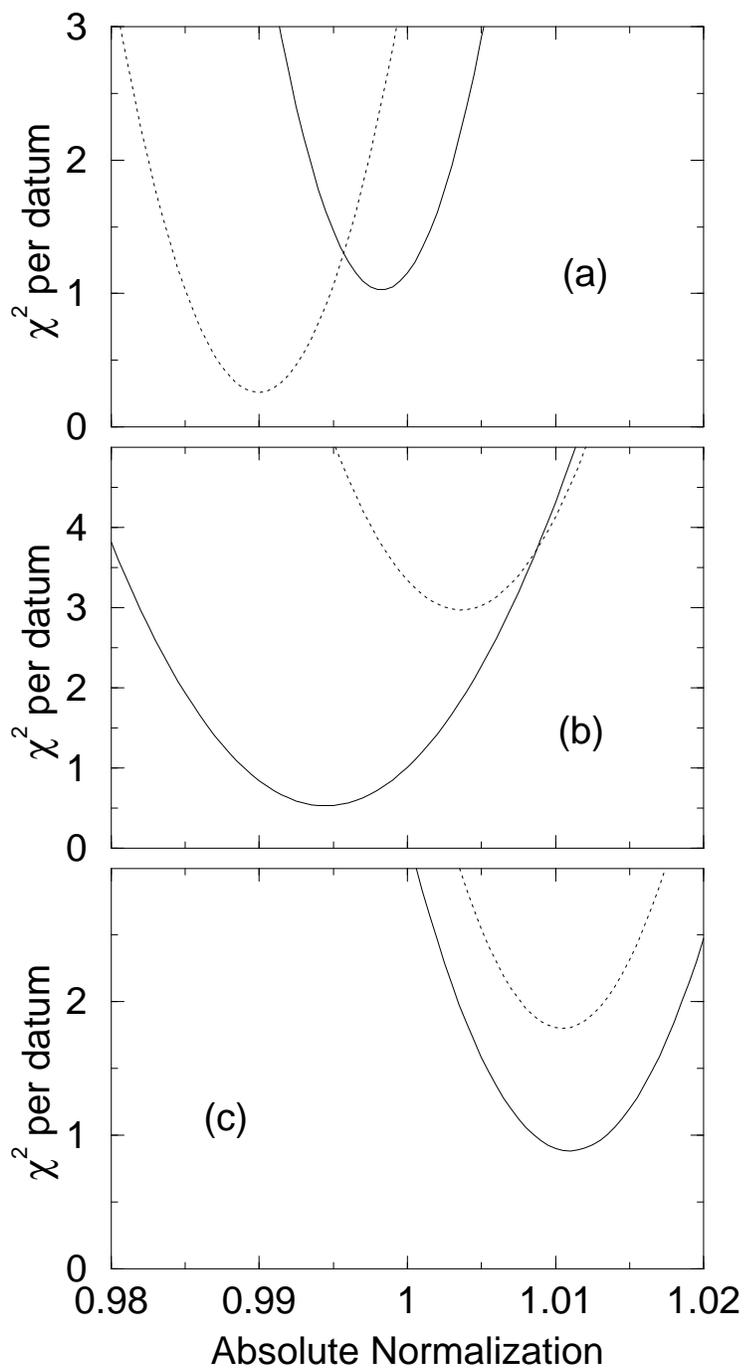
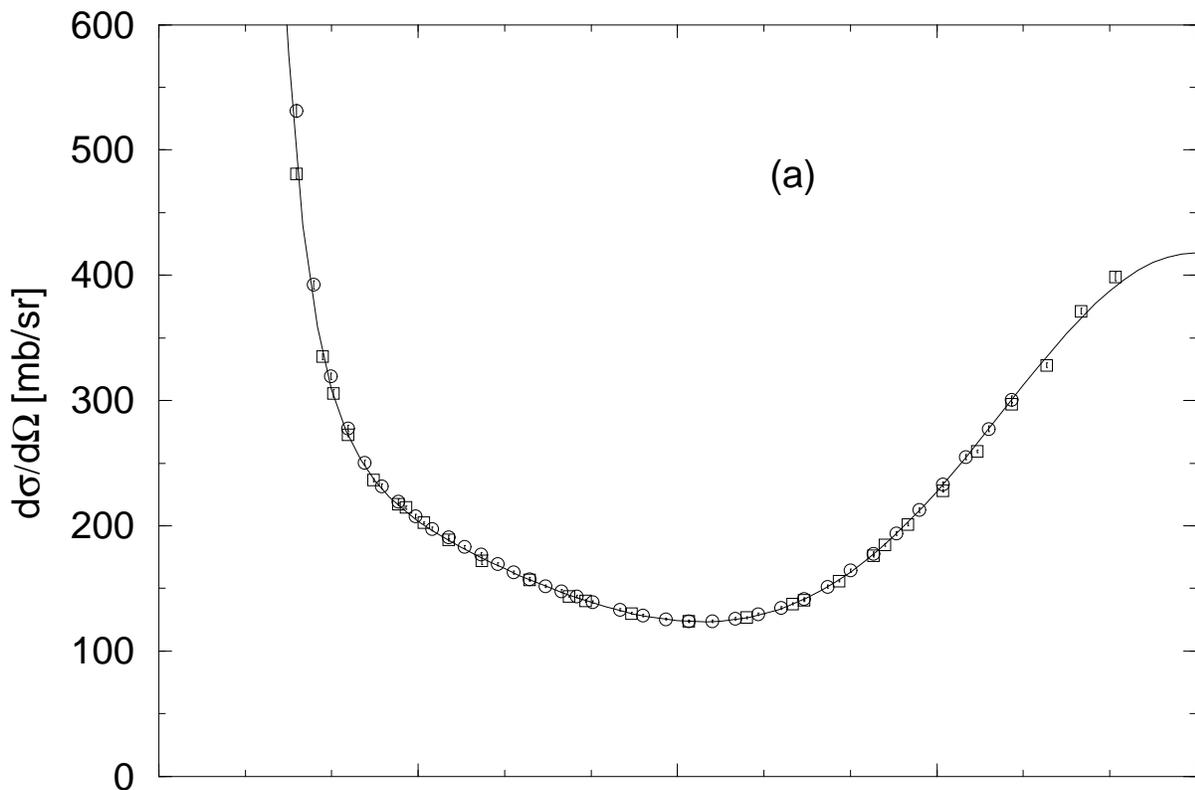
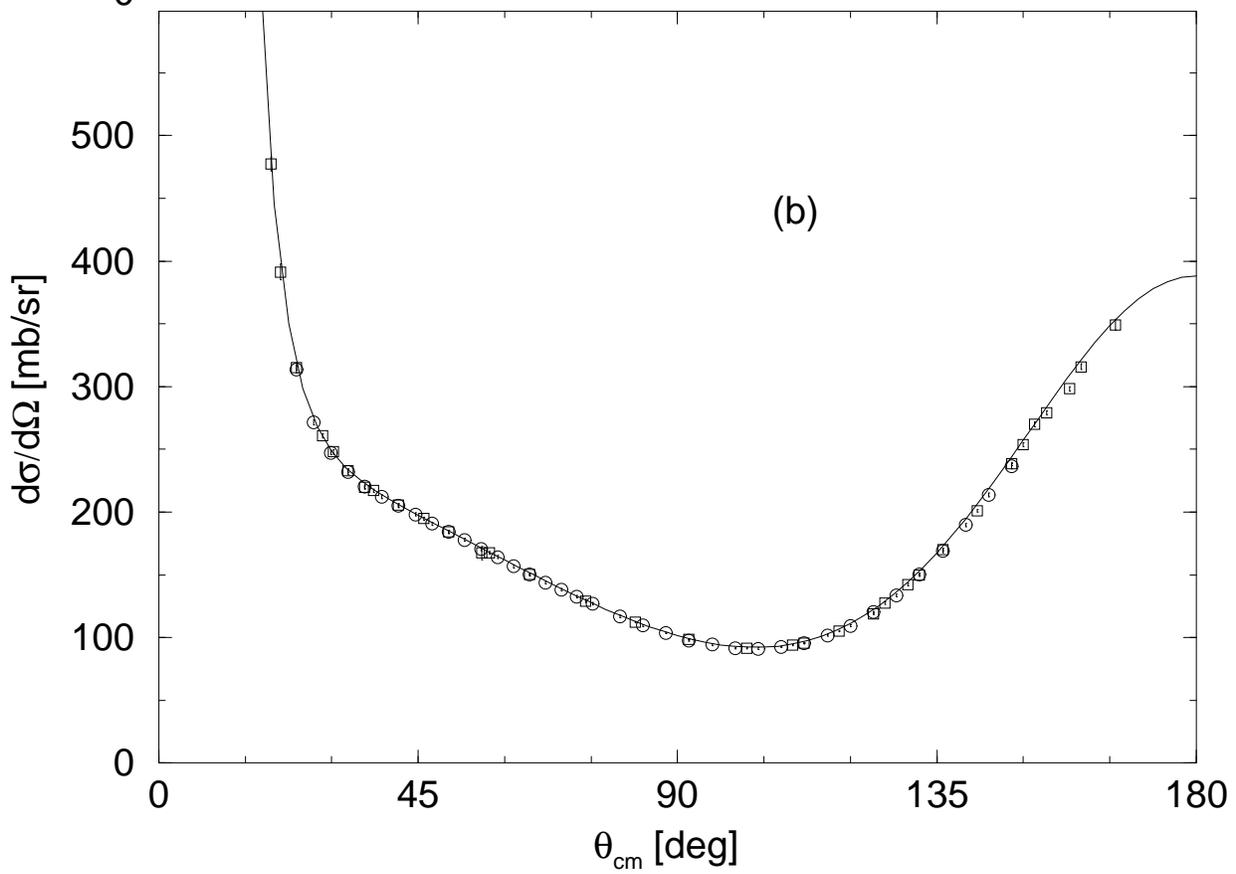
\end{document}